\documentclass[letterpaper]{article}
\usepackage{interspeech2010,amssymb,amsmath,epsfig}
\usepackage[safe]{tipa}
\usepackage{tipx,nicefrac,url}
\def\reg{{\rm\ooalign{\hfil
     \raise.07ex\hbox{\scriptsize R}\hfil\crcr\mathhexbox20D}}}

\title{Topological Considerations for Tuning and Fingering Stringed Instruments}

\makeatletter
\def\name#1{\gdef\@name{#1\\}}
\makeatother
\name{{\em Terry Allen, Camille Goudeseune}}
\address{Beckman Institute, University of Illinois at Urbana-Champaign \\
{\small \tt terry@guitarmath.com, cog@illinois.edu}}
\begin{document}
\maketitle
\begin{abstract}

We present a formal language for assigning pitches to strings for
fingered multi-string instruments, particularly the six-string guitar.
Given the instrument's tuning (the strings' open
pitches) and the compass of the fingers of the hand stopping the strings,
the formalism yields a framework for simultaneously optimizing three things:
the mapping of pitches to strings,
the choice of instrument tuning, and the key of the composition.
Final optimization relies on heuristics idiomatic to the tuning,
the particular musical style, and the performer's proficiency.

\end{abstract}

\section{Introduction} 

The `guitar fingering problem' is, less colloquially,
the finding of an optimal assignment of pitches to strings and frets.
Previous work on this problem
\cite{Radicioni, Radis04, Rutherford09, Sayegh89, Tuohy05, Tuohy06, TuohyDn}
takes as fixed three things: the composition, the musical key, and the guitar's tuning.
From these, a fingering is found that demands little technical dexterity,
or that at least obeys the mechanical limitations of the instrument.

We generalize this by fixing only the composition,
and then simultaneously optimizing over three aspects:
the fingering, the tuning, and the key (transposing from, say, G major to A major).

Historical motivation for exploring nonstandard tunings is found in the
work of guitarists such as
Muddy Waters, Chuck Berry, and Keith Richards~\cite{richards},
John Lennon~\cite{everett},
and especially Michael Hedges~\cite{forte,hedges}.
The search for optimality is also driven by the observation that some
genres such as blues or Gypsy Jazz \cite{williams} typically use only one
or two fingers at a time.  Strong constraints are thus
critical for reconstructing fingering from audio-only recordings.

\section{Definitions} 

A {\bf pitch} is the fundamental frequency of a sound with a harmonic spectrum,
such as that played by many musical instruments.
Pitch is measured in log Hz.
The logarithm conveniently lets us define an {\bf interval}, such as an octave or semitone,
as a {\em ratio} of pitches (2:1 or $\sqrt[12]{2}$:1, respectively).\footnote{This is
a musical interval.  In context, there will be no confusion between this
and `mathematical' intervals like $[0,1] \subset \mathbb{R}$.}

Pitch is continuous, but as we are discussing fretted stringed instruments,
we temper the octave into twelve equal semitones (equal temperament).
Then we represent an interval as not a ratio but an integer
counting the number of semitones between two pitches.
Thus, in the familiar way, 12 represents an octave, 7 a perfect fifth, 1 a semitone.
If we fix a reference pitch,
then we can similarly represent any {\em pitch} as an integer,
namely the integer representing the interval from the reference to the pitch in question.
In the context of a particular tuning of an instrument,
the reference pitch is often that of the lowest string, played unfretted (see below).

A {\bf string} is defined by what it can do:
sound a pitch from a contiguous range
bounded at both ends by the fingerboard's finite length.
At any given moment, a string sounds either no pitch or exactly one pitch.
When a string sounds its lowest pitch,
we call it unfretted or {\bf open}.

An {\bf instrument} means a fretted multi-string musical instrument such as a guitar, arpeggione, or lute.
We concentrate on the six-string guitar because it is widely played,
often retuned, and used in many musical styles.

A {\bf fret} constrains a string's playable pitches to a discrete subset.
We assume that adjacent frets are one semitone apart,
and that all strings have the same frets (a rectangular fingerboard).\footnote{Bending
or `whamming' a string to sound a pitch outside equal temperament
is an issue separate to fingering, except for the detail that
an open string cannot be bent.}
We assume that strings are fretted by the four fingers of the left hand
(not the thumb, although some advanced styles permit the thumb to fret
the lowest string).  One finger can fret several adjacent strings.

An instrument's {\bf tuning} is the set of pitches of its open strings.
A {\bf tuning vector} is that set, sorted by rising pitch.
Thus, the number of strings determines the size of the tuning vector.
{\bf Retuning} means a change of tuning, whether or not this is to or
from an instrument's standard tuning (scordatura).
We ignore unusual `out of order' tunings such as Congolese mi-compos\'e,
standard tuning with the d string up an octave~\cite{Stewart}.
We also ignore tunings where several strings share the same pitch,
such as variations on Lou Reed's `Ostrich' (D-D-D-D-d-d).
In particular, we assume that as one moves across the fingerboard,
open pitch strictly rises.
Common examples for the guitar include
standard tuning (E-A-d-g-b-e$'$) and Open~G (D-G-d-g-b-d$'$);
hundreds more exist~\cite{wa,weissman,wikitunings}.\footnote{We
spell examples of tuning vectors with the older Helmholtz notation,
not scientific pitch notation (E$_2$-A$_2$-D$_3$-G$_3$-B$_3$-E$_4$)~\cite{young},
because the former is more legible in the few octaves used by the guitar.
We also spell such vectors with hyphen-separated pitch names,
instead of with mathematically orthodox tuple notation.}

A {\bf chord} is a set of pitches sounding simultaneously.
(These pitches need not have begun simultaneously.)
We abuse the term's traditional meaning by letting it contain fewer than
three pitches, what musicians call an interval or a unison.
This is because traditional harmonic analysis concerns us less than
the constraints on which strings can play which pitches.

A {\bf composition} is a finite sequence of chords of finite duration.
We concentrate our attention on the left hand, ignoring
right-hand techniques like plucking and damping.
A {\bf fingering} for a composition is a mapping from its pitches
to (string, fret) pairs.

A {\bf pitch value} is a pitch constrained to an integer (semitone) value.
A {\bf pitch class} is one of the twelve traditional names C, C$\sharp$, D, ..., B.
A {\bf musical key}, such as E major, is a set of seven
distinct pitch classes that forms a major or minor scale.
One of these pitch classes is special, called the {\bf tonic}.
A pitch's {\bf scale position} is its distance above the next lowest tonic pitch
of a musical key, measured in semitones.
Thus a scale position must lie in $[0, 12)$.

\section{Tuning Vectors}

Consider a tuning vector $(x_1, ..., x_6)$.
Because we start counting from the lowest string, the ${x_i}$ are
increasing.\footnote{Guitarists start counting from the highest string,
the one that is easiest to reach.  We do the opposite to simplify our notation,
and to obey the convention that a tuning's spelling
begins with the lowest string.}
We then define the corresponding {\bf length-5 tuning vector} as
\begin{equation*}
% \label{c5}
(C_1, ..., C_5)\ =\ (x_2-x_1,\; x_3-x_2,\; ...,\; x_6-x_5).
\end{equation*}
Because no two open strings share a pitch, the ${x_i}$ are {\em strictly} increasing:
$x_i \gneqq x_{i-1}$.  Thus each $C_i > 0$.
For convenience of summation in eqs.~\eqref{stringchange}, \eqref{boolean} and \eqref{redun}, we also define $C_0 = 0$.

One 5-vector (of intervals) corresponds to many 6-vectors (of pitches).
For example, (5,5,5,4,5), abbreviated as 55545, corresponds to both
E-A-d-g-b-e$'$ and D-G-c-f-a-d$'$).\footnote{Values of 4 and 5 are common
for guitar tunings, because of how wide the frets are spaced compared to the hand's size.}
Converting a 5-vector to a 6-vector, by choosing a pitch for the lowest string,
is what we call augmenting or {\bf evaluating} a 5-vector at a pitch.
For example, evaluating 75545 at D yields Drop-D tuning, D-A-d-g-b-e$'$.

For a given tuning,
let $PV_i(f)$ be the pitch sounded by string $i$ when fingered at fret $f$,
for $i$ from 1 to 6,
for $f$ from 0 to 24 (lowest pitch to highest, in both cases).
As a reference for this tuning, set $PV_1(0) = 0$.

Since frets are spaced one semitone apart,
\begin{equation*}
PV_i(f) = f + PV_i(0).
\end{equation*}

% Let $C_{i,j}$ be the interval between open strings $i$ and $j$, for
% arbitrary integers $1 \le i < j \le 6$.  (So, recalling eq.~\eqref{c5},
% $C_i~=~C_{i,i+1}$.)  We compute this by adding the intervals between
% consecutive strings:
% \begin{equation*}
% C_{i,j} = \sum_{k=i}^{j-1} C_k.
% \end{equation*}

Combining these yields the String Changing Equation
\begin{equation}
\label{stringchange}
PV_i(f) = f + \sum_{k=0}^{i-1} C_k
\end{equation}
for $1 \le i \le 6$ and $0 \le f \le 24$.
This equation characterizes the relationship between four concepts:
pitch value $PV$,
string $i$,
fret $f$,
and tuning vector $(C_1, ..., C_5)$.
This notation derives from work by Allen, formalized by Gilles and Jennings~\cite{Gilles}.

\section{Tones}

For notational convenience, we call $(i,f)~\in~\mathbb{Z}^2$
a {\bf guitar position}, or just a position.
As before, $1 \le i \le 6$ indicates string, and $0 \le f \le 24$ indicates fret.
This lets us define a {\bf tone} as an ordered triple:
(pitch value, scale position, guitar position).

Given a particular tuning and musical key,
a guitar can play some tones but not others.
For example, in standard tuning in E major, playing the lowest open string yields
the tone (0,~0,~(1,~0)): pitch value~0, scale position~0, (string~1, fret~0).
But the tone (0,~0,~(2,~0)) cannot be played: string~2, fret~0 has a different
pitch value and scale position than string~1, fret~0.
Playing the highest open string in the context of C major yields
the tone (24,~4,~(6,~0)): 24 semitones higher, 4 above C, sixth string, zeroth fret.

\subsection{The Vector Space of Tones}

If we momentarily relax some constraints on pitch value, scale position,
and the two elements of guitar position, then
$\mathbb{Z}~\times~\mathbb{Z}_{12}~\times~\mathbb{Z}^2$
describes the set of all possible tones.\footnote{We
abbreviate $\mathbb{Z}/12\mathbb{Z}$ as $\mathbb{Z}_{12}$, since there is no confusion
with the latter's other meanings in number theory.}
We still constrain scale position to lie in $[0,12)$.

If vector addition is componentwise, and (real) scalar multiplication is
multiplication on each element of the vector, then this set is not quite
a vector space over the field $\mathbb{R}$,
as it lacks closure under scalar multiplication.  For example,
$(8, 7, 6, 5)$ is in $\mathbb{Z} \times \mathbb{Z}_{12} \times \mathbb{Z}^2$,
but not $3.14 \times (8, 7, 6, 5)$.
To get a vector space, we must extend the $\mathbb{Z}$'s to $\mathbb{R}$'s.\footnote{To
restore closure, we could have constrained the scalars to be integers.
But then the scalars come from only the ring $\mathbb{Z}$ instead of the field $\mathbb{R}$,
leaving the space as only a module over $\mathbb{Z}$, {\em i.e.,} an abelian group.
This is too limited for our purposes.}
This demands continuous values for not only pitch, scale position,
and fret number, but also for string number.
(Imagine a Haken Continuum~\cite{haken} programmed to change pitch
along {\em both} horizontal axes.)

Then $\mathbb{Z}_{12}$ becomes $\mathbb{R}$ `mod 12.'
This is in fact a vector space over $\mathbb{R}$.
Intuitively, its vectors are directions on a clock face
and its scalars are the lengths of the clock's hands.
To prove this rigorously,
we define $\mathbb{R}_{12}$ as the set of length-1 vectors whose elements lie in $[0,12)$.
For vectors $[x], [y] \in \mathbb{R}_{12}$,
we define $[x] + [y] = [(x+y) - 12 \lfloor \frac{x+y}{12} \rfloor ]$.
For scalars $r~\in~\mathbb{R}$,
we define multiplication as $r~\cdot~[x] = [rx - 12 \lfloor \frac{rx}{12} \rfloor ]$.
The proofs of commutativity, distributivity, {\em etc.,} are then elementary.

As $\mathbb{R}$ and $\mathbb{R}^2$ are themselves vector spaces over $\mathbb{R}$ (the usual Euclidean ones),
we then conclude that the set $T$ of all tones is one too,
namely the direct sum $\mathbb{R} \oplus \mathbb{R}_{12} \oplus \mathbb{R}^2$.

\subsection{Movement of Tones}

As the set of all tones is a vector space,
we can imagine the subset of {\em playable} tones as a space within which we can move.
It then becomes useful to informally discuss `directions of motion' or `orthogonal vectors of tone movement.'
Also, instead of moving from one tone to another,
we can equivalently describe one playable tone `changing' to another,
as a side effect of changing guitar tuning, musical key, string, or fret.

The directions then correspond to the changing of one of the three
elements of a tone (pitch value, scale position, or guitar position).
Insofar as these three directions have an obvious unit value,
we can even call these orthogonal vectors orthonormal.

\subsubsection{Intuitive Structure of Playable Tones}\label{ortho}

Given a particular guitar tuning and a particular musical key,
the subset $S$ of playable tones has a certain structure relative
to $T$, the vector space of all possible tones.

We can imagine $S$ as a surface or manifold within $T$.
To motivate a derivation of its precise structure, consider
the example of tuning in fourths, E-A-d-g-c-f$'$ (55555 evaluated at E),
with the key of A major.
To intuit the local structure of $S$,
take a note somewhere in the middle of the fingerboard, such as (3, 4):
d string, fourth fret, sounding f$\sharp$.
Since f$\sharp$ is 9 semitones above A,
the corresponding tone in $S$ is (f$\sharp$,~9,~(3,~4)).
To find nearby members of $S$, change this tone elementwise:
\begin{enumerate}
\item If we change pitch value from f$\sharp$ to g, then 9 becomes 10, and (3, 4) becomes (3, 5).
More generally, as the finger moves along the string,
pitch value, scale position, and fret number move along a Euclidean line.
\item If we change the scale position from 9 to 10, the same thing happens.
\item If we change fret number from 4 to 5, the same thing happens.
\item If we change string from d to g, (3, 4) to (4, 4),
then f$\sharp$ becomes b and 9 becomes $9+5 \equiv 2 \pmod{12}$.
More generally, as the finger moves transversely across the strings,
pitch value and scale position (mod 12) move along a Euclidean line,
while fret value remains constant.
\end{enumerate}

Within this particular tuning and musical key,
scale position can be derived from pitch value:
a playable tone with pitch value $x$ must have scale position
$(s_0 + x) \pmod{12}$, where $s_0$ is the scale position of pitch value 0,
the lowest pitch of the lowest string.
This removes one dimension from $S$, leaving at most three.
But $S$ also needs at {\em least} three dimensions:
\begin{itemize}
\item From (1), pitch value is independent of string number.
\item From (1), fret number is independent of string number.
\item From (4), fret number is independent of pitch value.
\end{itemize}
Thus a local neighborhood (a topology) in $S$ has {\em exactly} three dimensions,
or three independent vectors.  We cannot call the vectors
suggested by (1) through (4) orthogonal,
because their dot product may be nonzero.  For example,
starting again from (f$\sharp$,~9,~(3,~4)), moving in `direction' (1) yields
(g,~10,~(3,~5));  in direction (4) yields (b,~2,~(4,~4)).  Computing the dot
product,
{\ninept
\begin{align*}
((\mathrm{g}, 10, 3, 5) - (\mathrm{f}\sharp, 9, 3, 4)) &\cdot
((\mathrm{b}, 2, 4, 4) - (\mathrm{f}\sharp, 9, 3, 4)) &=\\
(1, 10-9, 3-3, 5-4) &\cdot (5, 2-9+12, 4-3, 4-4) &=\\
(1,1,0,1) &\cdot (5,5,1,0) &=\\
1 \cdot 5 + 1 \cdot 5 &+ 0 \cdot 1 + 1 \cdot 0 = 10 \ne 0.
\end{align*}
}
Since $S$ might not contain the origin, it might not be
a proper vector subspace of $T$.  This dimensionality
argument still suggests that $S$ looks locally like a 3-dimensional
affine subspace of $T$.  But appearances deceive:
sections \ref{manifold} and \ref{ominimal} disprove this and find more fitting structures for $S$.

\subsubsection{Playable Tones are a PL Manifold}\label{manifold}

Let $S$ be the set of playable tones in $T$,
using the tuning vector 55555 from the example in section \ref{ortho}.
Viewed discretely, each element of $S$ corresponds to one guitar position.
Since a guitar position can take on 6 string values and 25 fret values,
$|S| = 6 \times 25 = 150$.
Viewed continuously, on the other hand, the finite fingerboard imposes bounds on $S$.
Guitar position must lie in $[1,6] \times [0,24]$.
This in turn bounds pitch values.
At the bottom, position (1,~0) has pitch value 0 by definition,
while at the top, eq.~\eqref{stringchange} says that
position (6,~24) has pitch value $PV_6(24) = 24 + (5+5+5+5+5) = 49$.

These bounds prevent $S$ from being closed under either linear or affine combinations,
so $S$ can be neither a subspace nor an affine subspace of $T$.

But when we induce a topology on $S$ from the usual topology
on $\mathbb{R}^3$, the one whose basis is the set of open spheres,
then $S$ becomes a manifold within $T$, a manifold with corners~\cite{joyce}.
``Any closed rectangle in $\mathbb{R}^n$ is a smooth $n$-manifold with corners''~\cite{lee}.
The closed rectangle here comes from the bounds themselves, $[1,6] \times [0,24] \times [0,49]$.
Also, since each string has comfortably more than 12 frets,
each string's set of pitches spans more than an octave.
Thus $S$ includes all possible values of scale position.

Having jumped from linear algebra to topology,
we can now dispense with the requirement of constant-interval tunings such as 44444 or 55555.
When the interval between adjacent strings is nonconstant,
$S$ merely relaxes from a manifold to a piecewise linear (PL)
manifold~\cite{moise1,moise2,moise3}.\footnote{As $S$ has fewer than four dimensions, it escapes Milnor's
counterexample to Hauptvermutung~\cite{milnor}.}

\subsubsection{The Set of all Playable Tones is O-Minimal on $\mathbb{R}$}\label{ominimal}

Recall that the set of tones $T$ is $\mathbb{R} \oplus \mathbb{R}_{12} \oplus \mathbb{R}^2$.
Consider $S$ as a subset of not $T$ but rather the `bigger' space $\mathbb{R}^4$.
Relabel as $x_1, ..., x_4$ the scalars that we have called pitch value, scale position, string, and fret.
Then we can define $S$ in terms of solution sets to inequalities that
are in the form of polynomials in the $x_i$.
We begin with the bounds on $S$,
expressed as two inequalities each for pitch value, scale position, string, and fret:
\begin{align*}
x_1 \ge 0 &\land x_1 \le 49 &\land \\
x_2 \ge 0 &\land x_2 < 12 &\land \\
x_3 \ge 1 &\land x_3 \le 6 &\land \\
x_4 \ge 0 &\land x_4 \le 24. &
\end{align*}
(Here, $a \ge b$ abbreviates the conjunction of the equation $a=b$ and
the inequality $a > b$.) Another Boolean expression specifies the correspondences
between pitch value $x_1$ and (string, fret) pair $(x_3,~x_4)$ found in eq.~\eqref{stringchange}:
\begin{align}
\label{boolean}
\bigvee_{i=1}^{6} \biggl(i-1 \le x_3 < i\biggr) \land \biggl(x_1-x_4 + \sum_{k=0}^{i-1} C_k = 0\biggr).
%    ((0 \le x_3 < 1) &\land (x_1 - x_4 &= 0))                    \quad &\lor 
% \\ ((1 \le x_3 < 2) &\land (x_1 - x_4 + C_1 &= 0))              \quad &\lor 
% \\ ((2 \le x_3 < 3) &\land (x_1 - x_4 + C_1 + C_2 &= 0))        \quad &\lor 
% \\ &...&                                                        \quad &\lor
% \\ ((5 \le x_3 < 6) &\land (x_1 - x_4 + C_1 + ... + C_5 &= 0)). \quad &
\end{align}
The conjunction of all these is a finite Boolean combination of sets of the forms
$\{(x_1, ..., x_4) : f(x_1, ..., x_4) > 0\}$ and
$\{(x_1, ..., x_4) : g(x_1, ..., x_4) = 0\}$,
where all the $f$'s and $g$'s are polynomials (all of degree 1, incidentally).
Hence, $S$ is a semialgebraic set.

Of the many interesting corollaries that follow from this result,
we note particularly that the set $\{S\}$ of all sets of playable
tones---all possible $S$'s, from all tunings, in all musical
keys---forms an o-minimal structure on $\mathbb{R}$.  This means that 
$\{S\}$ is `nice:' it is closed under common operations such as
finite unions and intersections, complements, and (via the Tarski-Seidenberg Theorem)
projection to lower dimensions~\cite{brumfiel,vddries}.

\subsubsection{Lines within Playable Tones}

The subset of $S$ that lies on one string has a powerful structural description,
namely a line.
If we call a tone's fret position $x$,
the pitch value of the tone's open string $b$,
and the tone's pitch value $y$,
then the line $y=mx+b$ relates $x$~to~$y$ within $S$, with the slope $m$ being 1.
Conjoining six of these equations, one per string, yields a vector equation
\begin{equation}
\mathbf{y} = \mathbf{x} + \mathbf{b}
\label{pitchpos}
\end{equation}
where
$\mathbf{x}$ is the vector of fret positions,
$\mathbf{b}$ is the guitar's tuning, and
$\mathbf{y}$ is the resulting vector of pitch values.

We call eq.~\eqref{pitchpos} the {\bf pitch-position relation}.
Particularly when written as $\mathbf{y} - \mathbf{x} - \mathbf{b} = \mathbf{0}$,
it describes the shape of $S$ for any tuning $\mathbf{b}$.
More concretely, if we re-tune from $\mathbf{b}_1$ to $\mathbf{b}_2$
but wish to leave pitches unchanged,
then eq.~\eqref{pitchpos} shows how the fingering's fret positions must
compensate by changing from $\mathbf{x}_1$ to $\mathbf{x}_2$:
\begin{align*}
\mathbf{y}_1 &= \mathbf{x}_1 + \mathbf{b}_1 & \text{(old tuning)} \\
\mathbf{y}_2 &= \mathbf{x}_2 + \mathbf{b}_2 & \text{(new tuning)} \\
\mathbf{y}_1 &= \mathbf{y}_2                & \text{(same pitches)} 
\end{align*}
Combining these and solving for $\mathbf{x}_2$, 
the new fret positions are $\mathbf{x}_2 = \mathbf{x}_1 + (\mathbf{b}_1 - \mathbf{b}_2)$.

In cryptography, such a difference of vectors, $\mathbf{b}_1 - \mathbf{b}_2$,
is called a Caesar shift or Caesar cipher.
Thus, we call this difference a {\bf cipher}.
A cipher `encodes' how a change of tuning causes a change of fingering.
Section \ref{Ciphers} treats this in detail.

% The tone's second element, scale position, is linear in $\mathbf{y}$ (and thus in $\mathbf{x}$).

\subsection{Three Directions of Tone Movement}

Recall that a local neighborhood in $S$ has three dimensions.
We can describe these three dimensions functionally:

\begin{itemize}
\item {\bf Transformation} means that a tone preserves its pitch value, but changes scale position.
\item {\bf Translation} means that a tone preserves its scale position, but changes pitch value.
\item {\bf Isomerization} means that a tone preserves both pitch value and scale position,
but changes guitar position.
\end{itemize}

One more term will become useful:
{\bf re-keying} means a change of musical key ({\em e.g.,} from E major to D major).

\subsubsection{Three Corollaries}

Three common applications arise when practically applying these directions.

\begin{figure}
\centerline{\epsfig{figure=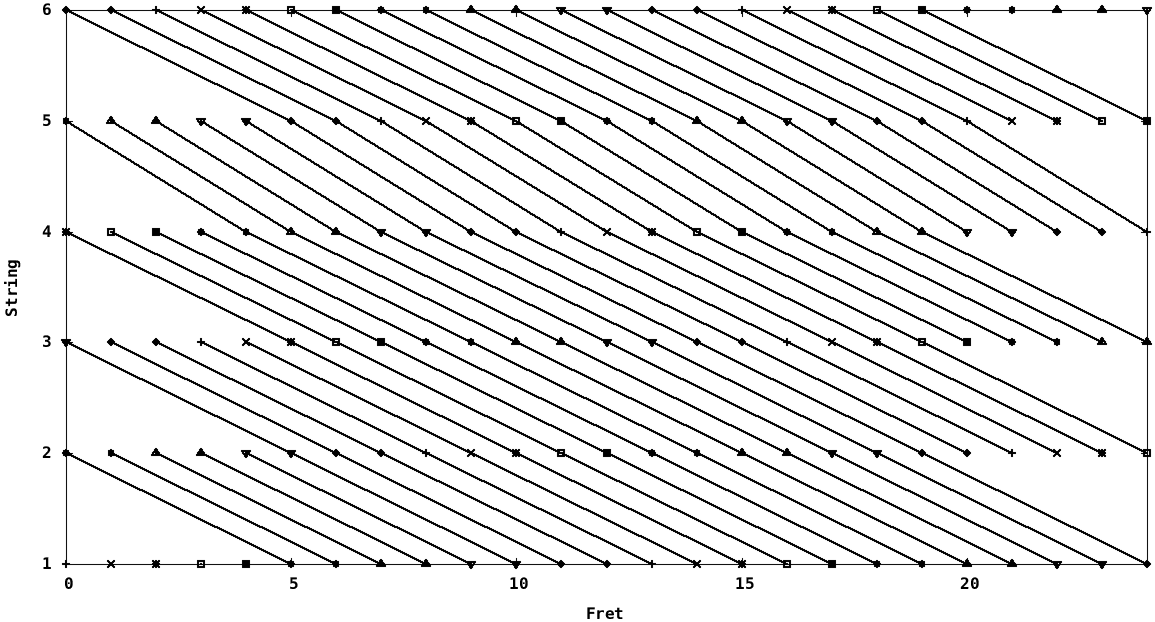, width=\linewidth}}
\caption{{\it Iso-pitch lines for standard tuning.  Fret pitch increases
from left to right;  string pitch increases from bottom (E) to top (e$'$).
The visibly different slope of the lines between strings 4 and 5 
corresponds to the `4' in standard tuning's 55545 tuning vector.}}
\label{fig:isopitch}
\end{figure}

\begin{itemize}

\item {\em Transformation usually implies retuning} (with a compensating
fret change, to leave pitch unchanged).  But transformation could
also mean that no retuning happened:  instead, the musical key changed,
landing the same pitch on a different scale position.
Transformation could even mean both retuning and re-keying.

\item {\em Translation implies re-keying.} For example,
in C major, pitch value G has scale position 7.
To remain at 7, but change pitch from G to A, the musical key
must change from C major to D major.
For this reason, we sometimes call translation transposition,
the everyday term for changing the key of a musical composition.

\item {\em Isomerization demands neither retuning nor re-keying.}
Because of the other two motions, transformation and translation,
without loss of generality we can constrain isomerization
to preserve both guitar tuning and musical key.
Isomerization really implies a change of string and fret,
leaving everything else constant.  It is motion along
{\em iso-pitch lines,} like contour lines on a topographic map
(fig.~\ref{fig:isopitch}).\footnote{Fig.~\ref{fig:isopitch} may mislead insofar as it is drawn in the plane.
Its ``45-degree'' appearance suggests that isomerization is a linear combination
of two other directions of motion of tones, as if $S$ had only two dimensions,
not the three proven in section \ref{ortho}.  But the axes of the figure
are string number and fret number, rather than directions of motion.  The figure
cannot represent anything about transformation or translation, because it
represents neither tuning vectors nor musical keys.  If it is to be interpreted
as a picture through which tones move, the picture is a severely restricted one.}

\end{itemize}

\subsubsection{Ciphers}\label{Ciphers}

A {\bf cipher} is a vector of 6 intervals.
A cipher can notate retuning, re-keying, or both.
In other words, given a tuning, a key, and a fingering,
if the tuning or key changes, then the cipher encodes how to
restore the original pitches by changing the fingering,
The word also alludes to how someone listening to a performance
must discover the `hidden' guitar tuning
{\em before} the actual fingering
(unravelling redundancy, section \ref{Redundancy}).

As a retuning example, to notate a change from
standard to Drop-D (E-A-D-g-b-e$'$ to D-A-d-g-b-e$'$),
the cipher is the elementwise difference between these
two tuning vectors (starting arbitrarily from C=0, so E=4, A=9, {\em etc.}):
{\ninept
\begin{equation*}
(4-2, 9-9, 14-14, 19-19, 23-23, 28-28) = (2,0,0,0,0,0).
\end{equation*}
}
To leave pitches unchanged after this retuning, the cipher indicates that
fret numbers increase by 2 on the first string, and remain unchanged on the
other strings.
More generally, the number of nonzero elements in a retuning cipher
indicates how many strings were retuned.  As a sign convention we let
positive elements correspond to strings retuned lower, so elements can be
directly added to fret numbers as a compensating offset to restore pitch.

On the other hand, as a re-keying example,
the cipher $(-2,-2,-2,-2,-2,-2)$ notates a key change
down two semitones, such as from E major to D major.\footnote{This same cipher could
also notate a {\em retuning up} two semitones, such as from E-A-d-g-b-e$'$ to F$\sharp$-B-e-a-c$\sharp'$-f$\sharp'$.
But in practice, a cipher with all elements equal indicates a change of key.}
Again, the cipher's elements work as fret offsets:  to play the same
composition in the lower key, fret numbers must decrease by 2 on all strings.

Like all vectors, ciphers add elementwise.  If we combined the previous two examples,
perhaps because the composition was in E major while Drop-D is more playable in D major,
then the resulting cipher would be $(0,-2,-2,-2,-2,-2)$.

Because a fret number $f$ is bounded ($0 \le f \le 24$),
adding an element of a cipher to $f$ can push it out of range.
When this happens, we replace that tone with another
that has the same pitch value (see section \ref{Redundancy}).
This is far more common with negative cipher elements pushing $f$ negative
than positive elements pushing $f > 24$.  When $f < 0$, the tone
often moves only to the next lower string (or lower still, until $f$ once
again reaches nonnegativity).  Avoiding excessively low strings
reduces stretching and jumping along the fingerboard (see section \ref{Playability}).
When precise voice-leading is not critical, the class of tone substitutions
can increase to include octave substitutions.  Less rigorous resolutions
of out-of-range fret numbers include replacing the tone with one whose
pitch value is a different element in the currently sounding chord
(``doubling a different note''), and even outright omission of the offending tone.

\section{Constraints}

When choosing a fingering, we first obey the mechanical limits
of the fingerboard and then consider more flexible desiderata.
The first of these we formalize in the term redundancy,
the second in playability.

These constraints apply to not just software, but also to intuitive
human searching for better fingering.  The ubiquity of chess-playing
software that is better than almost any human player has not dulled
the appeal of learning to play chess oneself.  Similarly, software that
finds optimal fingerings hardly makes manual search obsolete.  Indeed,
a guitarist might prefer an inferior fingering (that took longer to find)
to one found by computer but that demanded first typing in the notes,
since pleasure is to be found in the hunt itself.
This preference would in fact be forced for the many competent guitarists
who do not read music, or for compositions that are too improvisational
to be amenable to the task of data entry.

\subsection{Redundancy}\label{Redundancy}

Given a composition and a tuning, the most elementary choice in fingering is:
which (string, fret) pair should play a given pitch value?

We formalize this choice in a pitch value's redundancy,
the number of tones that have it as a first element.
Mathematically, the redundancy of a pitch value $p$ is
\begin{equation*}
|\lbrace \; i \; | \; 1 \le i \le 6\; \land PV_i(0) \le p \le PV_i(24) \rbrace|
\end{equation*}
or, applying eq.~\eqref{stringchange},
\begin{equation}
\label{redun}
\left|\left\lbrace i \; | \; 0 \le p - \sum_{k=0}^{i-1} C_k  \le 24 \right\rbrace\right|
\end{equation}
where $p$ is relative to the tuning $\lbrace C_k \rbrace$, that is,
pitch value zero corresponds to the lowest open string.
For example, in standard tuning, e and G have a redundancy of 2,
while E has a redundancy of 1.  (Of course,
moving a given pitch to a lower string demands a higher fret.)

\begin{figure}
\centerline{\epsfig{figure=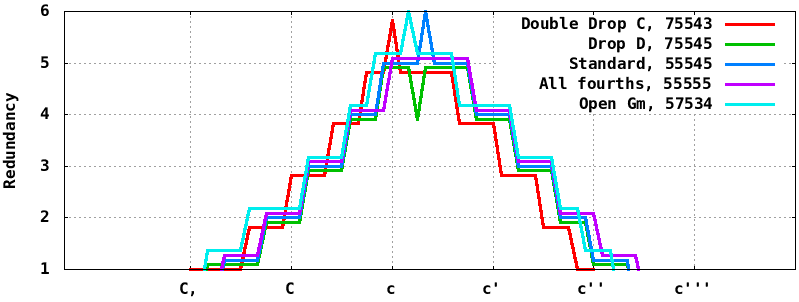, width=\linewidth}}
\caption{{\it Redundancy of pitch values in various tunings.}}
\label{fig:redundancy}
\end{figure}

If we shrink the intervals between adjacent strings of a tuning,
then pitch values will increase in redundancy (fig.~\ref{fig:redundancy}).
The inevitable compromise is that such a tuning has a smaller interval
between its lowest and highest strings, a smaller overall pitch range.
This means fewer iso-pitch lines in fig.~\ref{fig:isopitch}
(topographically speaking, more widely spaced contour lines,
that is, less steep terrain).  The varying pitch `elevations' of a composition
will then increase the width of both jumps and hand-stretches.

We empirically observe that the more playable tunings are those which
have redundancies slightly greater than one for a given composition's
pitch values.  Redundancies of 4 or more may cost too much in terms of
these jumps and stretches.

The outermost strings necessarily include some pitch values with redundancy 1.
For example, the lowest string includes pitch values too low to be played on any other string.
So from that redundancy-1 set of pitch values, only one at a time can sound.
Chords that include multiple members of that set cannot be played.
A similar constraint applies to the highest pitches on the highest string.
For example, in standard tuning, both the lowest-string and highest-string
redundancy-1 sets have 5 members.  The 5's come directly from the first and last digits
of that tuning vector, 55545.  The bottom left and top right corners of fig.~\ref{fig:isopitch}
render these as groups of 5 points, where each point is a degenerately short iso-pitch line.

\subsubsection{Capo}

For purposes of optimization, we may ignore capo without loss of generality.
A proof of this has two directions.  First, say a guitarist somehow finds an
acceptable tuning, key, and fingering for a composition.
If that happens to use a capo, and we remove the capo, then the same
fingering still works when moved down the appropriate number of frets:
the musical key will have merely been transposed.
Conversely, say someone wants to change musical keys to match a singer's
vocal range, without changing fingering.  A retuning may na\"\i vely
tighten all strings so far that the guitar neck deforms,
but a capo solves this problem more conventionally.

\begin{figure}
\centerline{\epsfig{figure=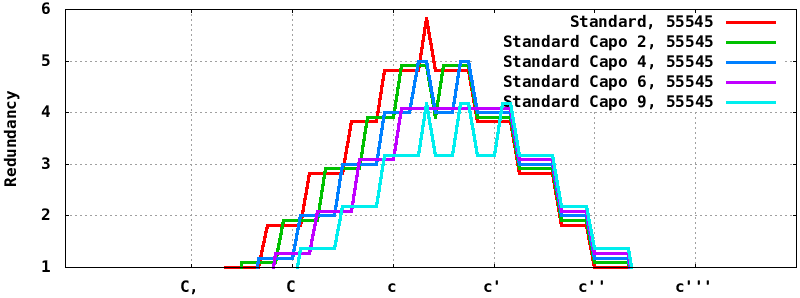, width=\linewidth}}
\caption{{\it As capo position increases, both redundancy and pitch range decrease.
This holds for all tunings, not just the standard tuning shown here.}}
\label{fig:capo}
\end{figure}

Fig.~\ref{fig:capo} shows how higher capos unsurprisingly reduce both
redundancy and pitch range.  So, at least for finding a fingering,
there is no harm in omitting the complicating extra degree of freedom
introduced by capo position.

\subsection{Playability}\label{Playability}

In the context of performing a composition, playability is our aim of
minimizing the demanded technical dexterity, the mechanical difficulty.
Among other things, we wish to avoid wide stretches and large fast jumps.
(This is just as true of keyboards as it is of fingerboards.)

Given a particular fingering, we assign a collection of costs to it as
suggested by previous work~\cite{Radicioni} \cite[p.~77]{Sayegh89} \cite[p.~501]{Tuohy05}.
To determine an overall cost, these costs are weighted to
suit a particular guitarist's preferences.  For example, one person
might dread jumps, while another jumps confidently but is limited in
chord choice by injury~\cite{williams} or arthritis.

\subsubsection{Stretches}

We prefer stretches that span fewer frets.
Thus, for any chord we define its {\bf span cost}.
This is 0 for a barre chord (plus open strings, if applicable).
It increases monotonically to 1 for a 5-fret span.
For larger spans it is $+\infty$, effectively forbidding such spans.
At the high end of the fingerboard, where frets are spaced more closely,
this may be less important, but a virtuoso comfortable in that stratosphere
needs little guidance anyways.

\subsubsection{Jumps}

We prefer shorter jumps.  Closely related to this, we prefer keeping the
hand at the low end of the fingerboard.  Empirically, this lets open
strings play more often;  it is also more familiar to beginning players.
For two consecutive chords, then, we define their {\bf jump cost}
as a monotonically increasing function of the jump size,
that is, the number of frets moved by the index finger.
(An absence of jumping yields a zero jump cost.)

\subsubsection{Open Strings}

We prefer open strings.
Thus, for any chord we define its {\bf fret cost}.
This is zero if no strings are fretted,
and increases monotonically to one
as the number of fretted strings increases to 6.
(Four fingers can fret more than four strings, as in barre chords.)

As a side effect of this preference for open strings,
a constraint between tunings and musical keys arises.
Given a tuning, the musical keys that are generally more playable are those whose
pitch classes are found in the tuning's open strings.
This is especially so for pitch classes with harmonically important roles
like the tonic, subdominant, and dominant:  three or four musical keys
satisfy this for most tunings.

As an elementary demonstration of this, guitar music performed in standard
tuning often uses the key of E major, where four open strings play
harmonically important pitches, namely E, A, b, and e$'$.
Conversely, standard tuning almost never uses the key of E$\flat$,
where the only diatonic open strings are d and g,
pitches with less important harmonic roles (leading tone and mediant).

When manually exploring fingerings and tunings, a `tuning-key path'
may happen:  as a new tuning is tried, a new key suggests itself,
which may in turn suggest another tuning, and so on.

We prefer slower jumps.  For two consecutive chords, we define their
{\bf speed cost} as the duration expected to perform the jump divided by
the duration allowed.  The expected duration is
predicted by Fitts's Law~\cite{MacKenzie}.
Consequently, jumps towards the low end of the fingerboard,
where frets are spaced more widely, are less costly.
This agrees with empirical observation.

\section{Notation}

Once an acceptable tuning, key, and fingering is found,
tablature notates this result better than a five-line staff.
Tablature directly notates fingering, and is thus better for evaluating playability.
Although staff notation is better suited to the tools of traditional music theory,
it needs awkward diacritical marks to map pitches to particular strings and frets,
and the guitar's tuning must be extraneously indicated.

Note that without a tuning vector, tablature is as meaningless as a staff lacking a clef and key signature.

% \eightpt
\bibliographystyle{IEEEtran}

\begin{thebibliography}{99}
\bibitem{wa} Allen, W., \emph{WA's encyclopedia of alternate guitar tunings.} \url{http://members.cox.net/waguitartunings/tunings.htm}, 2010.

\bibitem{brumfiel} Brumfiel, G., \emph{Partially Ordered Rings and Semi-Algebraic Geometry.} Cambridge Univ. Press, 1979.

\bibitem{vddries} van den Dries, L., \emph{Tame topology and o-minimal structures.} Cambridge Univ. Press, 1998.

\bibitem{everett} Everett, W., \emph{The Beatles as musicians: the Quarry Men through Rubber Soul.} Oxford Univ. Press, p.~21, 2001.

\bibitem{forte} Forte, D., ``New directions in acoustic steel-string,'' in \emph{Guitar Player}, Feb. 1985.

\bibitem{Gilles} Gilles, F., and Jennings, D., private communication, 2011.

\bibitem{haken} Haken, L., Tellman, E., and Wolfe, P., ``An indiscrete music keyboard,''
in \emph{Computer Music J.} 22(1), pp.~30$-$48, 1998.

\bibitem{hedges} Hedges, M., and Stropes, J. \emph{Rhythm, sonority, silence.} Stropes, 1995.

\bibitem{joyce} Joyce, D., \emph{On manifolds with corners.} \url{http://arxiv.org/abs/0910.3518}, 2010.

\bibitem{lee} Lee, J., \emph{Introduction to Smooth Manifolds.} Springer, p.~364, 2002.

\bibitem{MacKenzie} MacKenzie, I.S., ``Fitts' law as a research and design tool in human-computer interaction,''
in \emph{Human-computer interaction}, pp.~91$-$139, 1992.

\bibitem{milnor} Milnor, J., ``Two complexes which are homeomorphic but combinatorially distinct,''
in \emph{Ann. Math.} 74(2), pp.~575$-$590, 1961.

\bibitem{moise1} Moise, E., ``Affine structures in 3-manifolds: V,'' in \emph{Ann. Math.} 56, pp.~96$-$114, 1952.

\bibitem{moise2} Moise, E., ``Affine structures in 3-manifolds: VIII,'' in \emph{Ann. Math.} 59, pp.~159$-$170, 1954.

\bibitem{moise3} Moise, E., \emph{Geometric topology in dimensions 2 and 3.} Springer, 1977.

\bibitem{Radicioni} Radicioni, D., and Lombardo, V., ``Guitar fingering for music performance,''
in \emph{Proc. Int. Computer Music Conf.} 2005, pp.~527$-$530, Barcelona, 2005.

\bibitem{Radis04} Radisavljevic, A., and Driessen, P., ``Path Difference Learning for Guitar
Fingering Problem,'' in \emph{Proc. Int. Computer Music Conf.} 2004, Miami, 2004.

\bibitem{richards} Richards, K. and Fox, J., \emph{Life.}  Little, Brown, \& Co., 2010.

\bibitem{Rutherford09} Rutherford, N., \emph{FINGAR, a genetic algorithm approach to producing
playable guitar tablature with fingering instructions.}
Undergraduate project dissertation, Dept. of Computer Sci., Univ. of Sheffield, 2009.

\bibitem{Sayegh89} Sayegh, S., ``Fingering for string instruments with the optimum path paradigm,''
in \emph{Computer Music J.} 13(3), pp.~76$-$84, 1989.

\bibitem{Stewart} Stewart, G., \emph{Rumba on the river: a history of the popular music of the two Congos.} Verso, p.~34, 2004.

\bibitem{Tuohy05} Tuohy, D., and Potter, W., ``A genetic algorithm for the automatic generation of playable guitar tablature,''
in \emph{Proc. Int. Computer Music Conf.} 2005, pp.~499$-$502, Barcelona, 2005.

\bibitem{Tuohy06} Tuohy, D., and Potter, W., ``Generating guitar tablature with LHF notation via DGA and ANN,''
in \emph{Proc. IEA/AIE}, pp.~244$-$253, Annecy, France, 2006.

\bibitem{TuohyDn} Tuohy, D., \emph{Creating tablature and arranging music for guitar with genetic algorithms and artificial neural networks.} M.S. Thesis, Univ. of Georgia, 2006.

\bibitem{weissman} Weissman, D., \emph{Guitar tunings: a comprehensive guide.} Routledge, 2006.

\bibitem{wikitunings} Wikipedia, \emph{Guitar tunings.} \url{http://en.wikipedia.org/wiki/Guitar_tunings}, 2011.

\bibitem{williams} Williams, D., and Potokar, T., ``Django's Hand,''
in \emph{British Medical J.} 339(7735), pp.~1427$-$8, 2009.

\bibitem{young} Young, R., ``Terminology for logarithmic frequency units,''
in \emph{J. Acoust. Soc. Am.} 11(1), p.~134$-$9, 1939.

\end{thebibliography}

\end{document}